\documentclass{llncs}

\usepackage{xspace,amsmath,url}
\usepackage{arydshln}

\usepackage{graphics}
\usepackage{epstopdf}
\usepackage{epsfig}

\usepackage{booktabs}

\usepackage{array}
\newcolumntype{L}[1]{>{\raggedright\let\newline\\\arraybackslash\hspace{0pt}}m{#1}}
\newcolumntype{C}[1]{>{\centering\let\newline\\\arraybackslash\hspace{0pt}}m{#1}}
\newcolumntype{R}[1]{>{\raggedleft\let\newline\\\arraybackslash\hspace{0pt}}m{#1}}

\newcommand{\exclude}[1]{}

\begin{document}

\title{Rank and select: Another lesson learned}

\author{Szymon Grabowski
and
Marcin Raniszewski
}

\institute{
	  Lodz University of Technology, Institute of Applied Computer Science,\\
	  Al.\ Politechniki 11, 90--924 {\L}\'od\'z, Poland \\
	  \email{\{sgrabow|mranisz\}@kis.p.lodz.pl}
}

\maketitle

\begin{abstract}
Rank and select queries on bitmaps are essential building bricks of 
many compressed data structures, including 
text indexes, membership and range supporting spatial data structures, 
compressed graphs, and more.
Theoretically considered yet in 1980s, these primitives have also been 
a subject of vivid research concerning their practical incarnations 
in the last decade.
We present a few novel rank/select variants, focusing mostly on speed, 
obtaining competitive space-time results in the compressed setting.
Our findings can be summarized as follows: $(i)$ no single rank/select solution 
works best on any kind of data (ours are optimized for 
concatenated bit arrays obtained from wavelet trees for real text datasets), 
$(ii)$ it pays to 
efficiently 
handle blocks consisting of all 0 or all 1 bits,
$(iii)$ compressed select does not have to be significantly slower than 
compressed rank at a comparable memory use.
\end{abstract}

\section{Introduction}
\noindent 
Rank and select are essential building bricks of many compressed 
data structures, and text indexes in particular.
In their most frequently used binary incarnation, they can be defined 
as follows:
given a bit-vector $B[0 \ldots n-1]$, 
$rank_b(B, i)$ returns the number of occurrences of symbol $b$ 
in the prefix $B[0 \ldots i]$ 
and $select_b(B, i)$ returns the position of the $i$-th occurrence 
of symbol $b$ in $B$, where $b \in \{0, 1\}$.

Note that $rank_0(B, i) = i - rank_1(B, i)$, hence it is enough 
to directly support the rank only for one binary symbol (e.g., 1).
There is no similar relation binding the values of $select_0(B, i)$ 
and $select_1(B, i)$.

It is known for at least two decades~\cite{Jac1989,Clark1996,Munro1996} that 
these operations can be performed in constant time, 
using the extra space of $o(n)$ bits.
Raman et al.~\cite{RamanRR02} showed how to compress the vector $B$ to 
$n H_0(B) + o(n)$ bits, where $H_0(B)$ is the order-0 entropy of $B$, 
and still support rank and select in constant time.

Much research has been dedicated to construct rank and select solutions, 
both compressed and non-compressed, to answer the queries possibly fast 
in practice.
Especially in the compressed setting also the lower-order terms 
of the space matter, 
hence the practical questions involve two aspects: the query time and the 
space used by the data structure.
The next section briefly recalls
the history of practical solutions for rank 
and select, starting from the non-compressed ones.

\section{Related work}
\noindent 
The original rank solution, by Jacobson~\cite{Jac1989}, 
which needs $O(n\log\log n/\log n)$ bits of space\footnote{Logarithms 
of base 2 are used throughout the paper.}
apart from the original bit-vector,
performs three memory accesses (to one superblock counter 
and one block counter, plus a lookup into a table 
with precomputed popcount answers), which often translates into 
three cache misses, a significant penalty.
Gonz{\'a}lez et al.~\cite{GGMN05} proposed a scheme with 
one level of blocks followed by sequential scan.
In theoretical terms, this solution no longer obtains both constant time 
and sublinear extra space, yet it fares well practically and is very simple.

Vigna~\cite{V08} interleaved data from different levels to improve access locality.
Gog and Petri~\cite{GP13} carried this idea even further, 
interleaving the precomputed rank values and the bit-vector data. 
The sequential scans over small blocks of data are performed with 
an efficient 64-bit hardware popcount instruction ({\em popcnt}),
available in Intel and AMD processors since 2008.
In the manner of the Gonz{\'a}lez et al. solution, 
Gog and Petri store only one level of precomputed ranks, yet 
data from two successive cache lines can be sometimes read in their scheme.
In a recent work, Grabowski et al.~\cite{GrabowskiRD2015} showed a solution 
with one cache miss in the worst case.
They achieve it with interleaving 64-bit precomputed rank fields 
with $512 - 64 = 448$ bits of data.
As 64 bits per rank is more than needed, part of this field stores 
popcount values for some prefixes of the following block, thus 
saving on the {\em popcnt} invocations.

K{\"{a}}rkk{\"{a}}inen et al.~\cite{KKP14} proposed a hybrid scheme 
for the compressed rank, where the bit-vector is divided into blocks 
and each block is encoded separately, choosing one of a few different methods, 
depending on its ``local'' performance.
This general approach was implemented in a version with three encodings:
no compression (i.e., the block kept verbatim), storing the positions of the 
minority bits (zeros or ones, whichever have fewer occurrences in the block), 
and run-length encoding for runs of zeros and ones.
To make the data structure even more compact, blocks are grouped into 
superblocks. 
Thanks to it, the blocks' header data store ranks and offsets to 
the beginnings of the encoded block bodies with respect to the beginning 
of the superblock rather than the beginning of the whole structure.
Only the rank operation is supported, yet the authors mention briefly 
a possibility to extend their scheme in order to support selects too.

As it can be implied from the literature, an efficient select 
is harder to design than an efficient rank, even in the non-compressed variant.
Clark~\cite{Clark1996} was the first to show a constant-time select 
with $3n/ \lceil \log\log n \rceil + O(n^{1/2} \log n \log\log n)$ 
bits of extra space.
The solution is relatively complicated and needs at least 60\% space overhead.
Gonz{\'a}lez et al.~\cite{GGMN05} noticed that implementing 
select with binary search over a rank structure is often superior
(in spite of having $O(\log n)$ time complexity), 
both in execution times and the space overhead.
Yet, for large inputs or for low densities of set bits (assuming 
that we focus on the $select_1$ query), Clark's solution dominates.
More recently, Gog and Petri~\cite{GP13} presented a practical implementation 
of the Clark select idea, reducing its worst-case space overhead 
to less than 29\%.

Okanohara and Sadakane~\cite{OkanoharaS07} were the first to consider 
practical {\em compressed} implementations of rank/select structures 
and they introduced four novel r/s dictionaries 
(each of which was based on different ideas), reaching different 
space/time tradeoffs in theory and in practice.
For example, they offer to answer rank or select queries 
in a few tenths of a microsecond (on a 3.4\,GHz Intel Xeon) 
spending about 25\% of extra space for densely (50\%) populated bit-vectors.
Vigna~\cite{V08} obtained similar times, yet with about twice smaller 
space overhead.
Navarro and Providel~\cite{NPsea12} raised the bar even higher 
(or, should we say, lower?), reducing the space overhead to about 
10\% of the original bit-vector on top of the entropy, solving 
in this space both rank and select. 
Their rank queries are handled within about 0.4 $\mu$sec and 
select queries within 1 $\mu$sec.
They also show how to reuse sampling data between the rank and the select 
in a non-compressed scenario, 
with a benefit in space, which allows to answer these queries 
within around 0.2 $\mu$sec, using only 3\% of extra space on top of 
the plain bit-vector.

A unique approach was taken by Beskers and Fischer~\cite{BF14}, 
who focus on sequences with low {\em higher-order} entropies. 
Their solutions are likely to be competitive e.g. 
for representing wavelet trees for repetitive collections of
strings, like individual genomes of the same species.

\section{Our algorithms}

In the two following subsections we present our compressed rank and select 
variants.

\subsection{Compressed rank}
\label{sec:crank}
The input bit-vector $B$ is conceptually divided into blocks of $k$ bytes, 
and each run of $h$ successive blocks is grouped in a superblock.
For each superblock a fixed-size header is stored, having $h$ ranks of the 
prefix of $B$ up to the current block and $h$ offsets to the areas storing the 
successive blocks.
Popcounting over a block (or its prefix), with a 64-bit built-in instruction, 
is used in the final phase of the operation.
Several variants were implemented, depending on the header and 
block representations.
One important optimization is used in all the variants:
a block consisting entirely of zero or one bits is not stored, as, 
during the query, such a case is easily recognized from checking 
two adjacent rank values.
Such a block will be called a {\em mono-block}.
Our variants are presented in the successive paragraphs.

\paragraph*{\textbf{basic}}
We have uncompressed rank and offsets in the superblock, 
each stored on 4 bytes, and uncompressed bit data.
(Note that the notion of a superblock is artificial for this variant, 
but we keep it for consistency.)

\paragraph*{\textbf{bch (basic with compressed headers)}}
The first rank and the first offset in a superblock are stored on 4 bytes, 
while the following $h-1$ ranks are stored differentially 
with respect to the first rank, using 2 bytes each.
Additionally, we keep $h-1$ bits (rounded up to a multiple of 16 bits) 
to tell which blocks are mono-blocks.
If the query falls into a non-mono-block, knowing the first offset and 
checking how many mono-blocks in the superblock precede the block in question, 
we immediately obtain the desired offset.

\paragraph*{\textbf{mpe1 (mono-pair elimination v1)}}
The bit data are compressed in a simple manner. 
The input block of size $k$ bytes is divided into 2-byte chunks 
and the chunks consisting of all zeros (i.e., two bytes 0) 
or all ones (i.e., two bytes 255) are not stored in the block. 
Such chunks are called mono-pairs.
To made decoding possible, the block is prepended with 
two fields of $k/2$ bits each 
(both rounded up to the nearest multiple of 16 bits), 
where the bits from the first field denote if the successive 2-byte chunks 
are stored in the block or not, 
and the bits from the second field distinguish between mono-pairs with 
zeros and mono-pairs with ones.
(Note that this encoding is somewhat redundant, as e.g. $k/2$ {\em trits} 
would be enough, but we preferred convenience over 
striving for maximum compression.)
Some blocks are however not (well) compressible, that is, 
the number of mono-pairs in them is less than $k / 16$.
In such cases, compression is not applied to the block.
The header contains one verbatim rank and one verbatim offset value, each 
stored on 4 bytes, 
while the following $h-1$ ranks and $h-1$ offsets are stored differentially, 
with respect to the first rank/offset, using 2 bytes each.
One bit out of the 2 bytes per differentially encoded offset is spent for the 
flag informing if compression is applied to the block.

\paragraph*{\textbf{mpe2}}
This is a little refinement of the previous variant. 
We separately check if it pays to remove mono-pairs of zeros, 
and mono-pairs of ones from the block.
There are four possible cases: 
no compression, 
eliminate only mono-pairs of zeros, 
eliminate only mono-pairs of ones, 
eliminate all mono-pairs, 
hence the choice is marked with 2 bits 
(again, taken from the differentially encoded offsets).

\paragraph*{\textbf{mpe3}}
The rank variants involving block compression are slower than 
{\em basic} or {\em bch}, due to the necessity to 
decode a compressed block before popcounting. 
To alleviate this penalty, in the current variant we apply compression 
only to the blocks which are reduced to at most half of their 
original size (we have not tried to tune this criterion).
In this way, weakly compressible blocks are stored verbatim 
and querying them is fast.

\paragraph*{\textbf{cf (cache-friendly)}}
In all the variants above, answering the rank query boils down to 
either noticing that the query falls into a mono-block, where the answer 
is immediate, or not, when the data block has to be accessed and processed 
appropriately.
The latter case is slower, partly because it typically involves another 
cache miss.
Therefore, if the fraction of mono-blocks in a dataset is $0 \leq f \leq 1$, 
we may expect about $2 - f$ cache misses per rank query, on average.
The current variant aims to reduce the average number of cache misses.
Let us have $b$ blocks in the input, $m$ of which are mono-blocks; 
the value of $m$ is found in the first pass over the data 
in the construction phase.
We need another scan over the blocks, and let us denote the 
number of mono-blocks among the first $j$ blocks with $m_j$ 
(that is, $m = m_b$).
We stop after such $j$-th block that 
$m_j = (b-j)-(m_b-m_j)$.
This divides $B$ into its left (the first $j$ blocks) 
and right part (the remaining $b-j$ blocks), 
in such a way that the number of mono-blocks in the left part 
is equal 
to the number of non-mono-blocks in the right part.
The data layout is now like that.
The blocks from the left part are located sequentially, 
each prepended with the 4-byte rank of the prefix of the sequence.
The right part stores 8 bytes per block in a contiguous area: 
4 bytes are for the rank and the other 4 bytes are for the offset. 
These offset values point to the mono-blocks from the left part.
Assuming that the mono-blocks are uniformly distributed over the whole 
dataset, the expected number of cache misses 
(or non-contiguous memory accesses) per query is 
$f \times 1 + (1-f)(1-f) \times 1 + f(1-f) \times 2 = 1 + f - f^2 \leq 2 - f$, 
where the equality holds only for $f = 1$.

\subsection{Compressed select}
\label{sec:csel}

The proposed select variants are related to the rank ones.
On a high level, our solutions make use of two integer parameters, 
$\ell$ and $thr$, and the $sel_1(B, j)$ values for all $j$ being a multiple 
of $\ell$ are stored directly.
Moreover, we distinguish between sparse blocks, that is such for which 
$sel_1(B, (i+1)\ell) - sel_1(B, i\ell) > thr$, and dense blocks for which 
the shown condition is not fulfilled.
If a block is sparse, the select values for all $j$ inside it, 
i.e., such that $i\ell < j < (i+1)\ell$, are stored directly.
For dense blocks we store the bits 
$B[sel_1(B, i\ell)+1 \ldots sel_1(B, (i+1)\ell)-1]$, 
possibly in a compressed form.
Again, block headers consist of precomputed select values and offsets to block data, 
and blocks are grouped into superblocks.
The particular variants: {\em basic}, {\em bch}, {\em mpe1}, {\em mpe2} and 
{\em mpe3}, mimic the corresponding rank variants, with some exceptions.
The main difference is that the select values in superblocks are never compressed 
(as opposed to rank values in several rank variants).
Additionally, {\em select-bch} differs to its rank counterpart in having 
its offsets encoded differentially (on 2 bytes) with respect to the first offset 
in the superblock.
Such a representation for select is needed to handle varying-length blocks.
Note also that the differential offset encoding enforces some restrictions 
on the $\ell$ and $thr$ parameters.

\begin{figure}
\centerline{
\includegraphics[width=0.48\textwidth,scale=1.0]{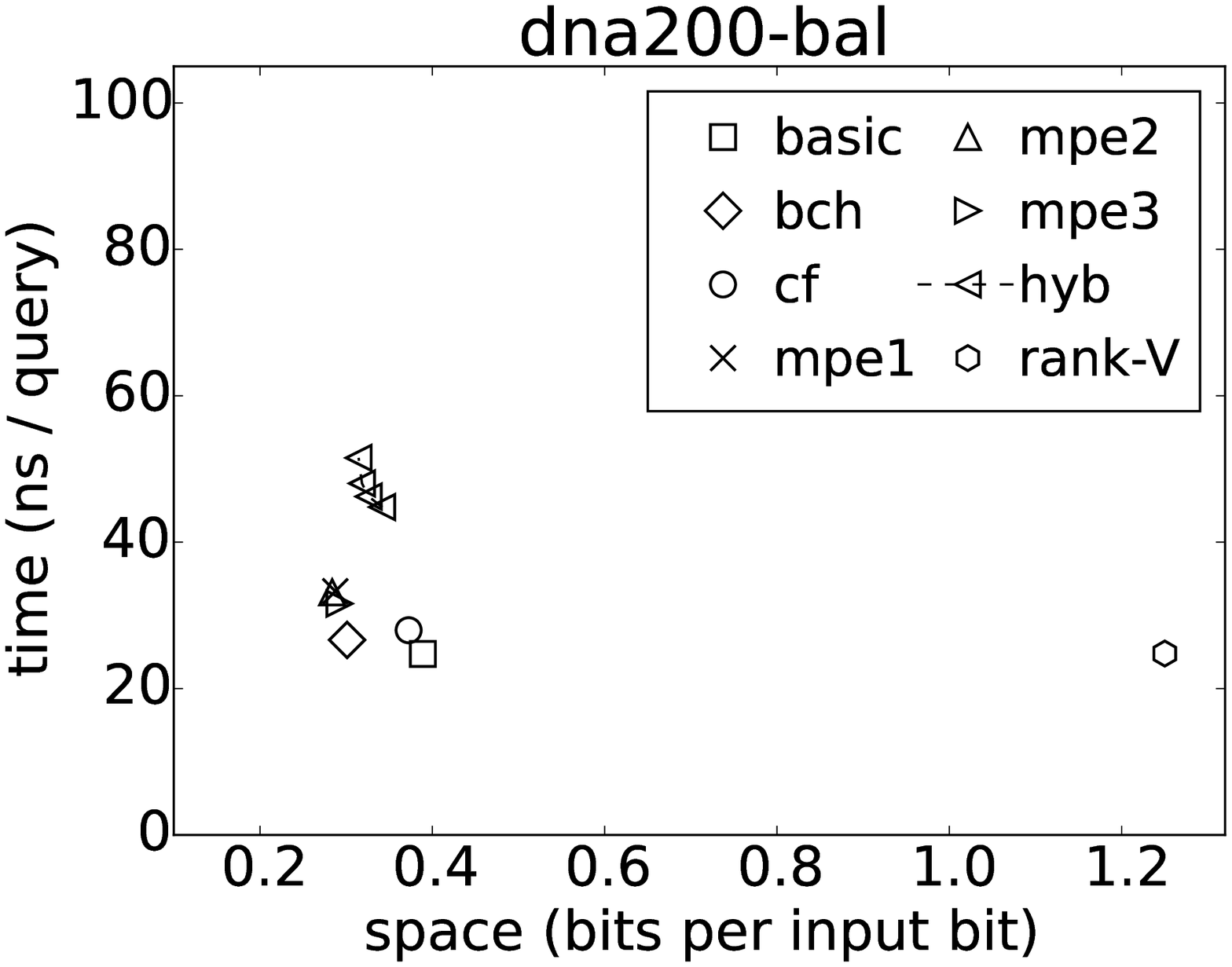}
\includegraphics[width=0.48\textwidth,scale=1.0]{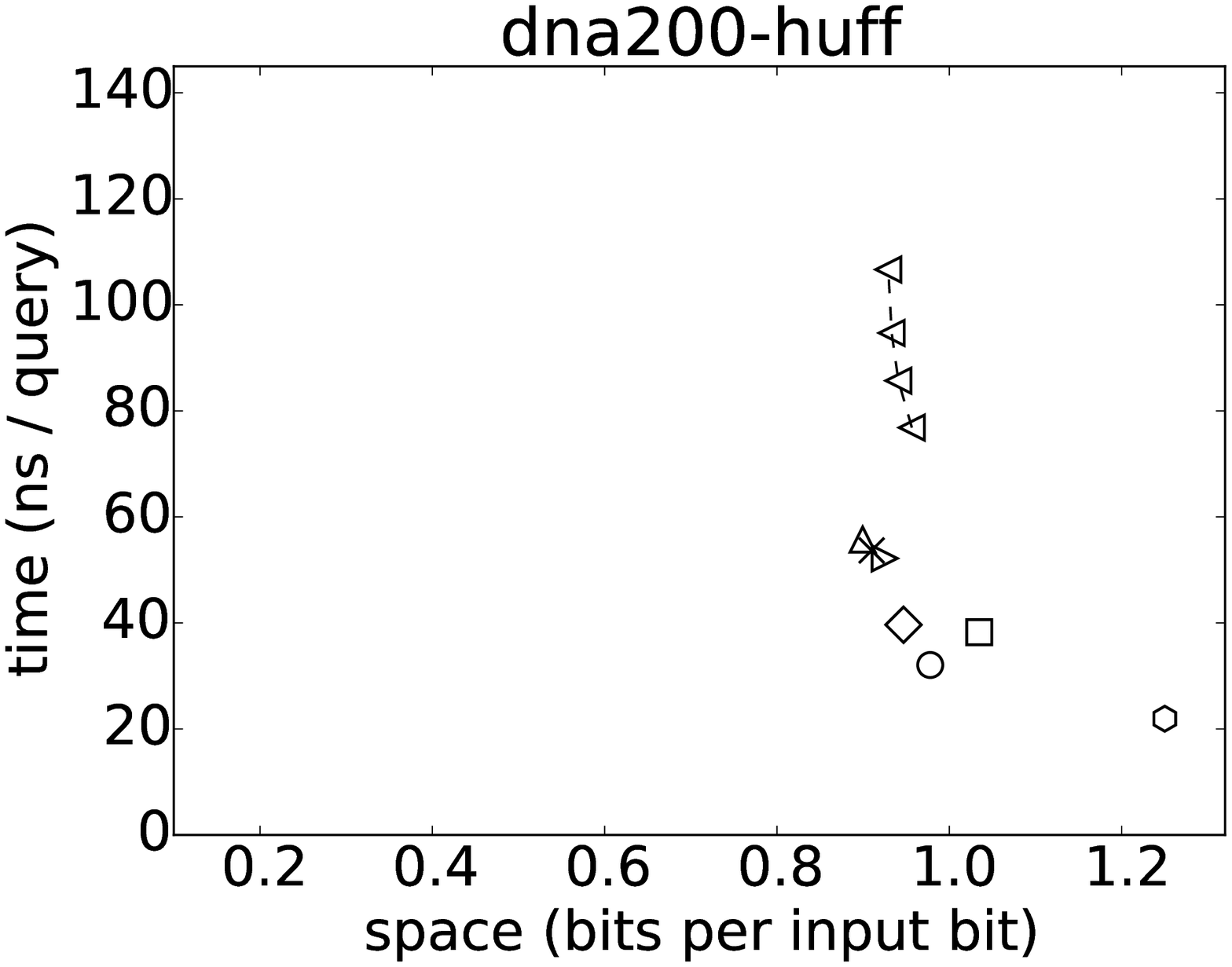}
}
\centerline{
\includegraphics[width=0.48\textwidth,scale=1.0]{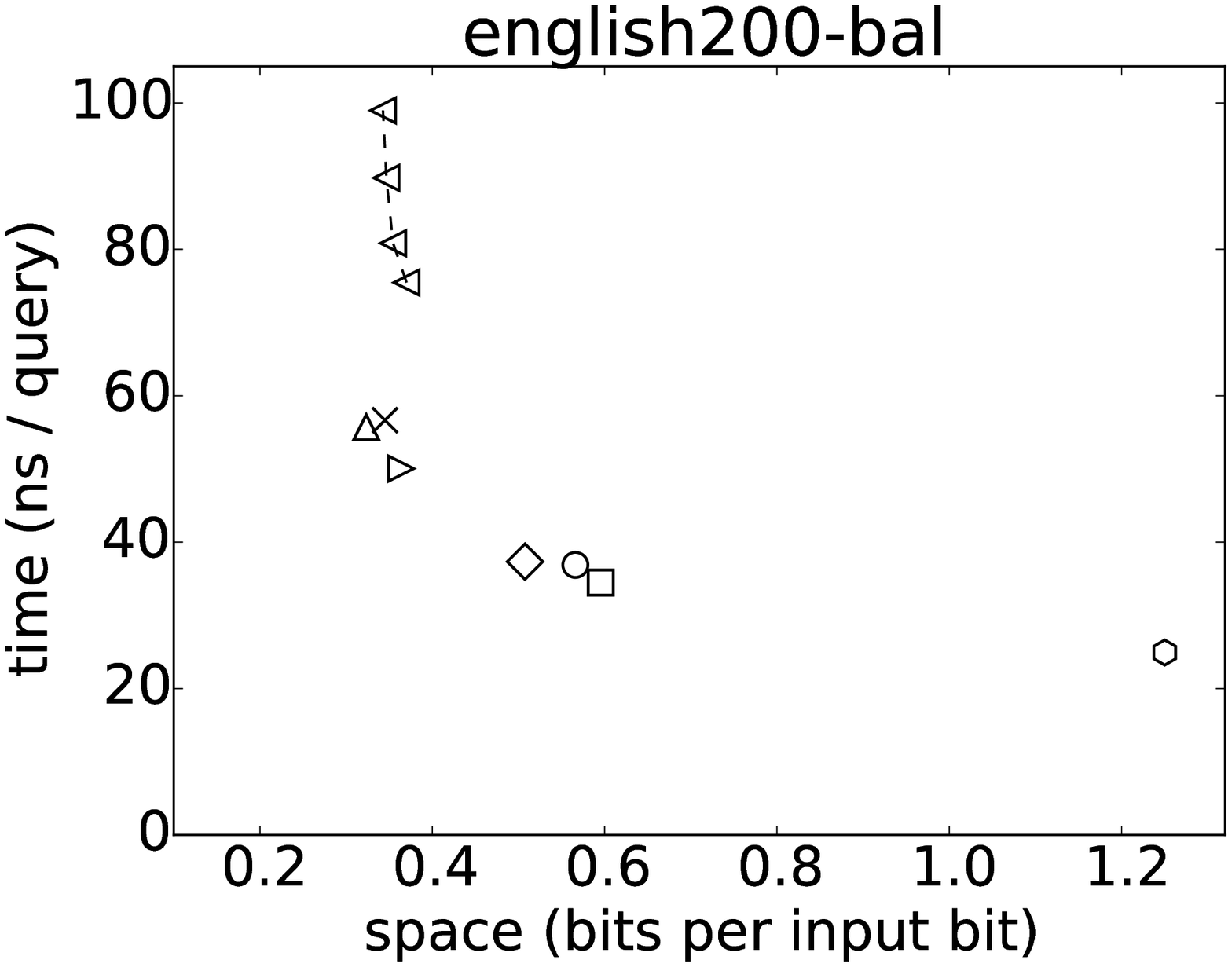}
\includegraphics[width=0.48\textwidth,scale=1.0]{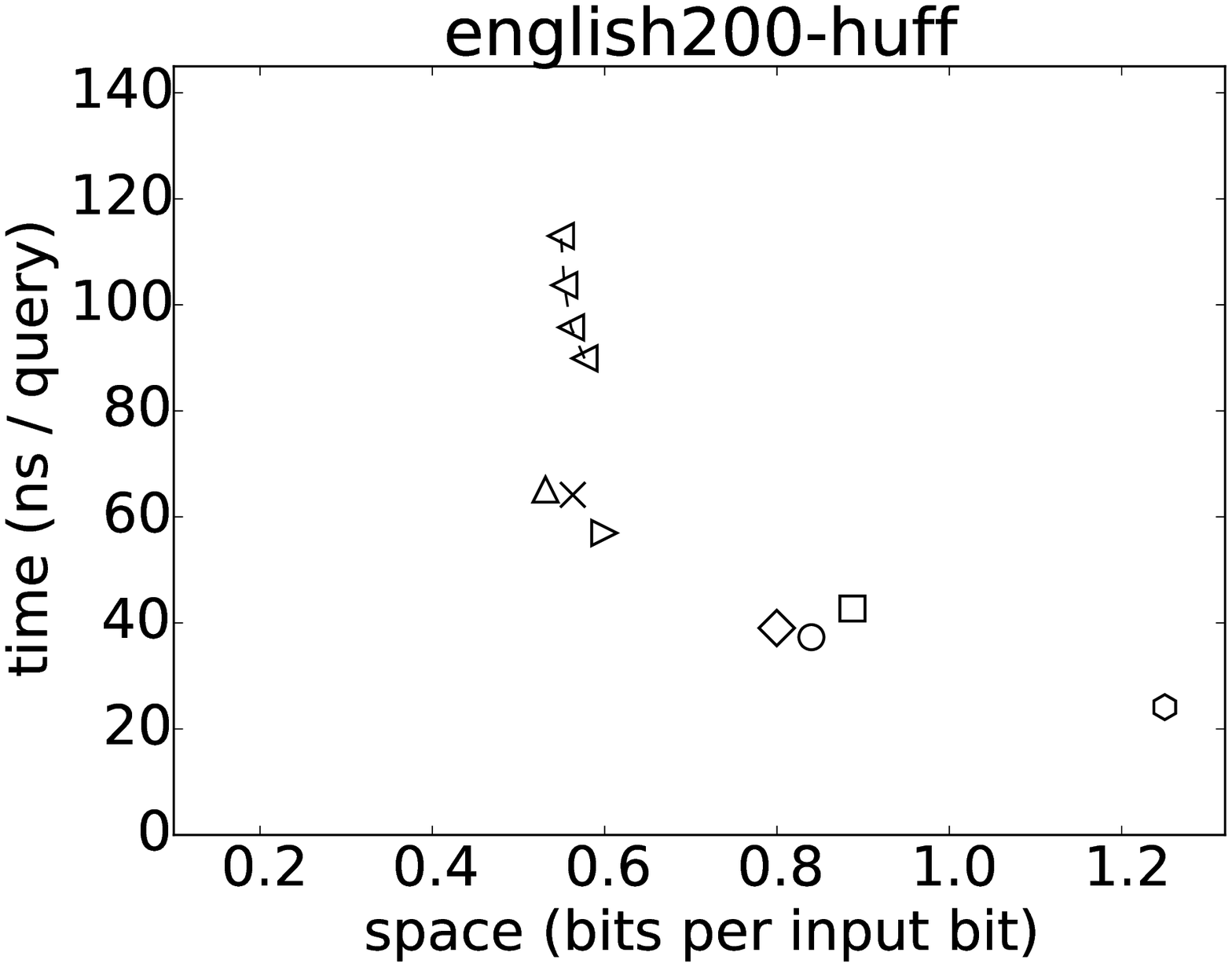}
}
\centerline{
\includegraphics[width=0.48\textwidth,scale=1.0]{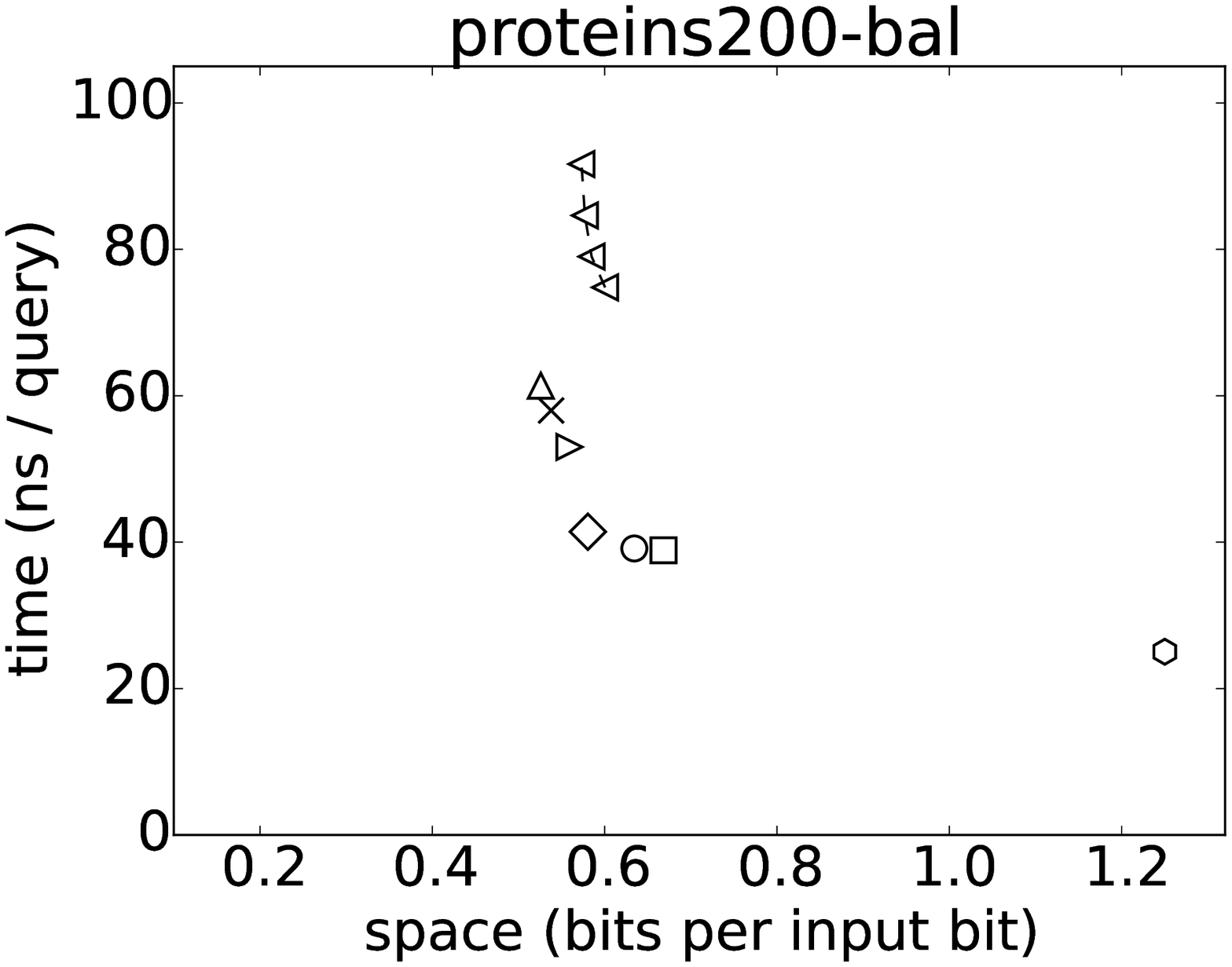}
\includegraphics[width=0.48\textwidth,scale=1.0]{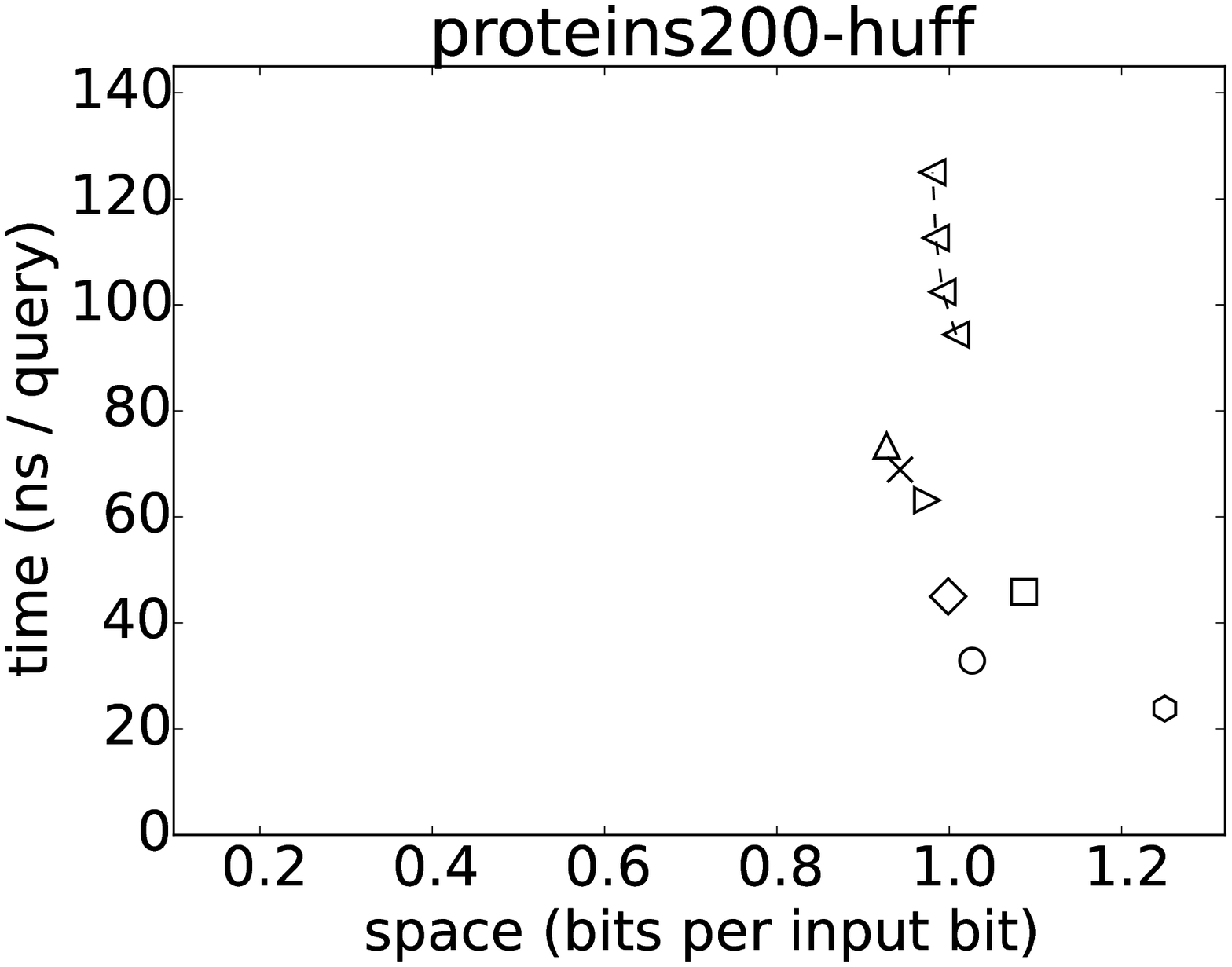}
}
\centerline{
\includegraphics[width=0.48\textwidth,scale=1.0]{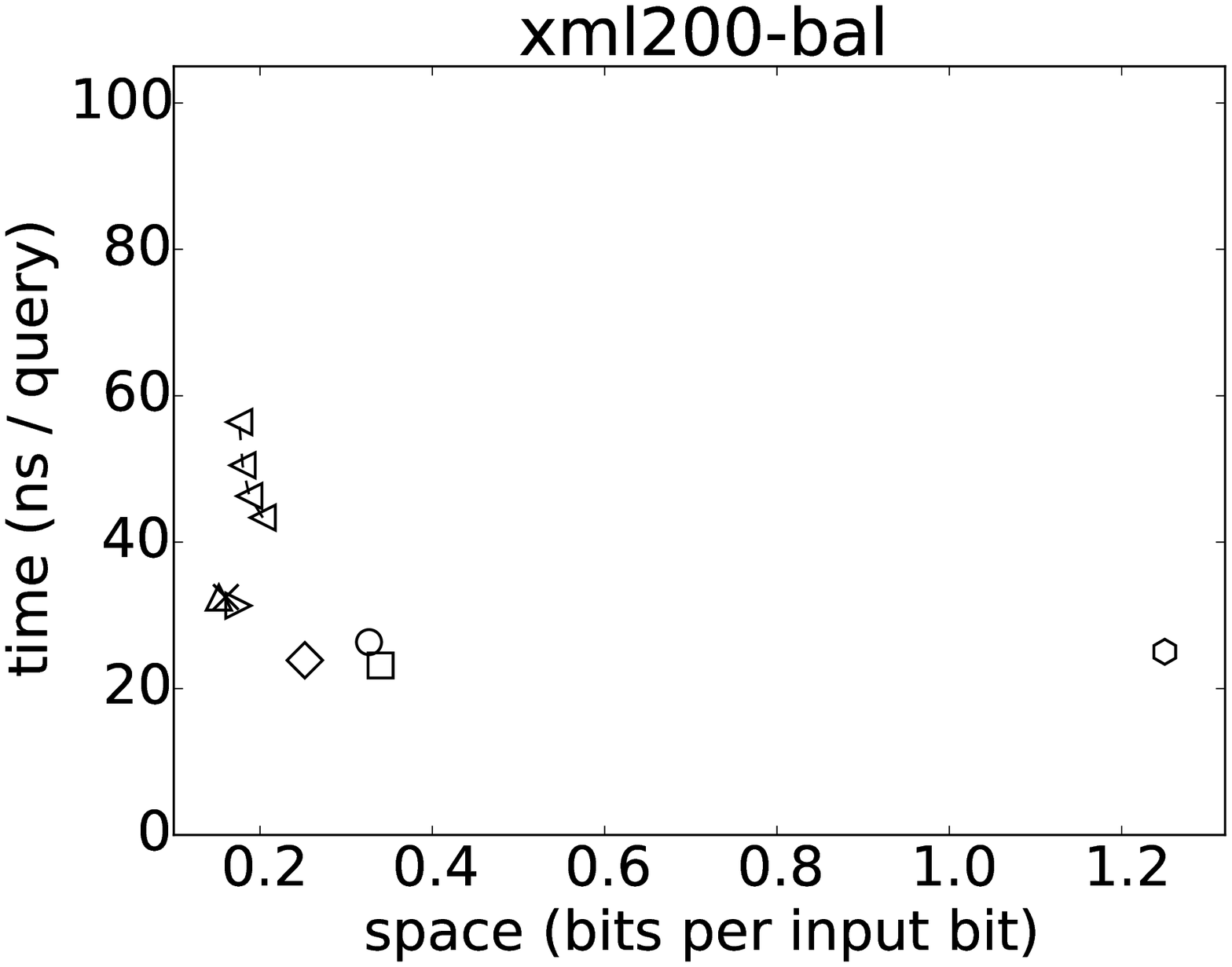}
\includegraphics[width=0.48\textwidth,scale=1.0]{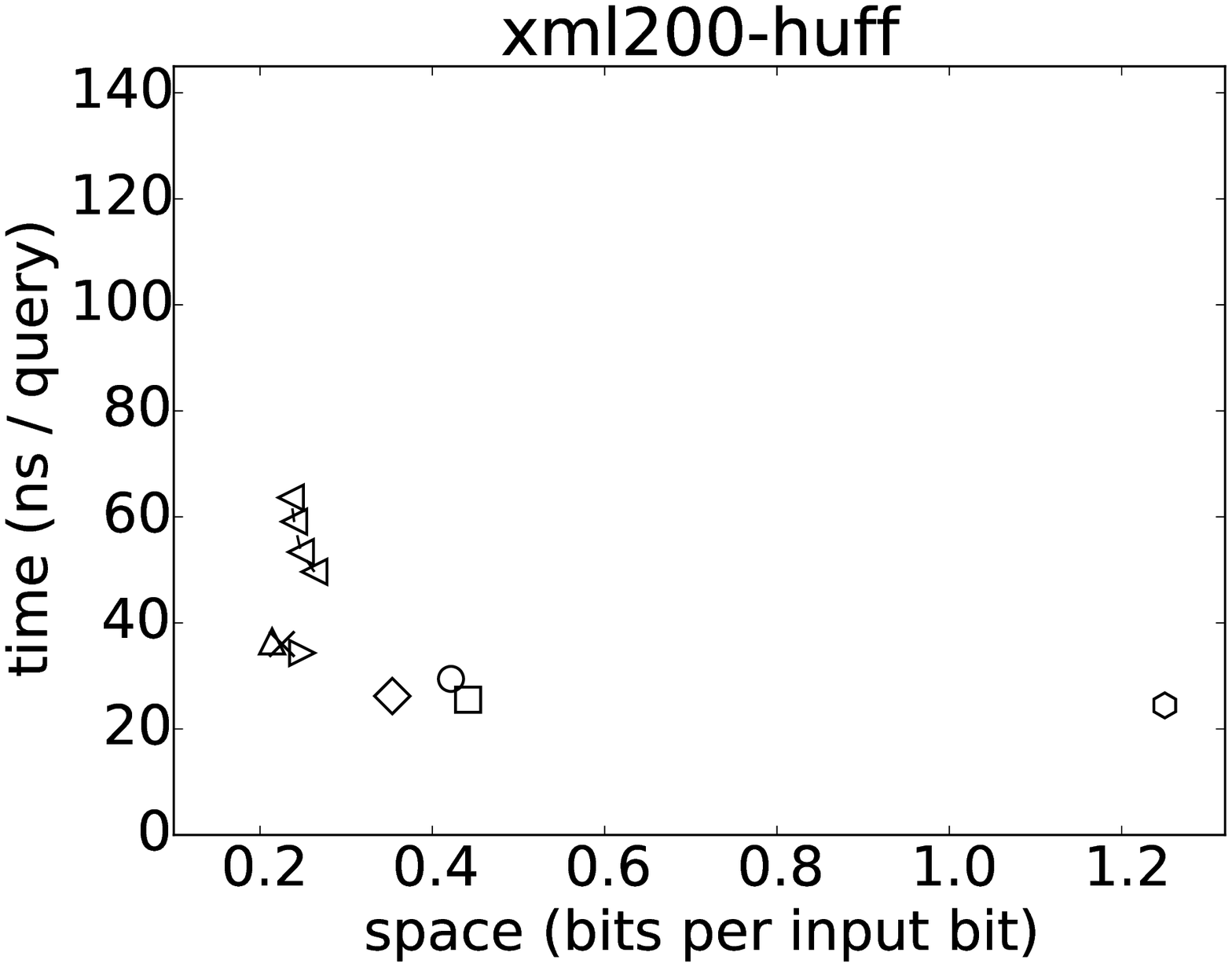}
}

\caption[Results]
{Rank query times (in ns), averaged over 100M random queries,
and used spaces for real data.
The datasets on the left: concatenated balanced WTs, 
on the right: concatenated Huffman-shaped WTs.
The {\em hyb} series parameters: 8, 16, 32 and 64.
}
\label{fig:cranks1}
\end{figure}

\section{Experimental results}
\label{sec:exp}
\noindent
All experiments were run on a machine equipped with 
a 6-core Intel i7 CPU 
(4930K) 
clocked at 3.4\,GHz, with 64\,GB of RAM, 
running Ubuntu 15.10 64-bit.
The RAM modules were $8 \times 8$\,GB DDR3-1600 with the timings 
11-11-11 (Kingston KVR16R11D4K4/64).
The CPU cache sizes were: 
$6 \times 32$\,KB (data) and $6 \times 32$\,KB (instructions) in the L1 level, 
$6 \times 256$\,KB in L2 
and 12\,MB in L3.
One CPU core was used for the computations.
All codes were written in C++ and compiled with 64-bit gcc 5.2.1,
with \texttt{-O3} and \texttt{-mpopcnt} options.
Our source codes 
are available at
\url{http://ranisz.iis.p.lodz.pl/indexes/ranks&selects/}.

The test datasets comprise concatenated bit-vectors from 
binary wavelet trees built over four 200\,MB Pizza~\&~Chili 
site\footnote{\url{http://pizzachili.dcc.uchile.cl/}} texts, 
in two versions, with standard (balanced) and Huffman-shaped WT layout.
Additionally, we used random bit-vectors with density of set bits 
equal to 0.05 or 0.2, respectively.

For comparisons, we took a number of rank and select variants
from the well-known sdsl library~\cite{gbmp2014sea}.
In particular, for the rank experiments we used:
\begin{itemize}
\item {\em hyb}, the hybrid rank from~\cite{KKP14}, with the 
superblock sizes of 8, 16, 32 and 64,
\item {\em rank-V}, a variant from~\cite{V08} (called {\em rank9} therein),
\item our variants: {\em basic}, {\em bch}, {\em mpe1}, {\em mpe2}, {\em mpe3} 
and {\em cf}, as described in Section~\ref{sec:crank}.
\end{itemize}
For the select experiments we took:
\begin{itemize}
\item Clark's algorithm~\cite{Clark1996} in an implementation 
proposed in~\cite{GP13}, called {\em sel-mcl}, 
\item {\em sel-RRR}~\cite{NPsea12}, called {\em sel-$R^3$K} in~\cite{GP13}, 
where $K$ is the block size, 
\item our variants: {\em basic}, {\em bch}, {\em mpe1}, {\em mpe2} and {\em mpe3}, 
as described in Section~\ref{sec:csel}.
\end{itemize}
The block size in the rank variants {\em basic}, {\em bch} and {\em cf}
was set to $k = 64$ and to $k = 128$ in the remaining ones.
The number of blocks per superblock, $h$, was set to 32 for {\em bch} 
and to 16 for the other rank variants and to all select variants.

From Fig.~\ref{fig:cranks1} we can see that 
our ranks {\em mpe1}, {\em mpe2} and {\em mpe3} 
offer similar (or slightly better) compression as {\em hyb} 
at a significantly higher speed.
If even more speed is preferred, then {\em bch} or {\em cf} is the method of choice.
Another Pareto-optimal variant is (in most cases) {\em basic}.
Not surprisingly, the only non-compressed solution in the comparison, 
{\em rank-V}, is the fastest, yet the gap between it and the other top 
contenders is often more striking in space than in time.

The implementation of {\em cf} is slightly more complicated (e.g., 
it has the extra, non-predictable, comparison at the start, to check 
if the parameter points to the left or to the right part of the bit-vector), 
and this is why it cannot beat {\em basic} in speed for the concatenated 
balanced WTs.
Table~\ref{table:cmisses} shows how the fraction of mono-blocks $f$ 
affects the average number of memory accesses per query and the 
total query times in the rank variants {\em basic} and {\em cf}.
The gap in the average number of memory accesses gets large for most 
Huffman-shaped WTs and there {\em cf} takes the lead in speed.

\begin{table}
\caption{The impact of the fraction of mono-blocks $f$ on 
the average number of memory accesses per query and the total query times 
in the rank variants {\em basic} and {\em cf}.
The block size is 
64 bytes.
The numbers of memory accesses are calculated from the formulas on $f$ 
given in Section~\ref{sec:crank}, in the paragraph on the {\em cf} variant.
The times are expressed in nanoseconds.}
\label{table:cmisses}
\setlength{\tabcolsep}{0.45em}
\begin{tabular}{L{2.45cm} R{1.85cm} C{1.5cm} C{1.5cm} C{1.5cm} C{1.5cm}}
\toprule
dataset	   & fraction of   & basic,   & cf,      & basic, & cf,  \\
           & mono-bl. [\%] & accesses & accesses & time   & time \\
\midrule
dna200-bal       & 73.60 & 1.2640 & 1.1943 & 24.73 & 27.97 \\
english200-bal   & 52.95 & 1.4705 & 1.2491 & 34.47 & 36.89 \\
proteins200-bal  & 45.66 & 1.5434 & 1.2481 & 38.89 & 39.14 \\
xml200-bal       & 78.50 & 1.2150 & 1.1688 & 23.16 & 26.34 \\
dna200-huff      &  9.03 & 1.9097 & 1.0822 & 38.25 & 32.06 \\
english200-huff  & 23.74 & 1.7626 & 1.1811 & 42.72 & 37.31 \\
proteins200-huff &  3.84 & 1.9616 & 1.0370 & 45.86 & 32.87 \\
xml200-huff      & 68.35 & 1.3165 & 1.2163 & 25.51 & 29.44 \\
\bottomrule
\end{tabular}
\end{table}

\begin{figure}
\centerline{
\includegraphics[width=0.48\textwidth,scale=1.0]{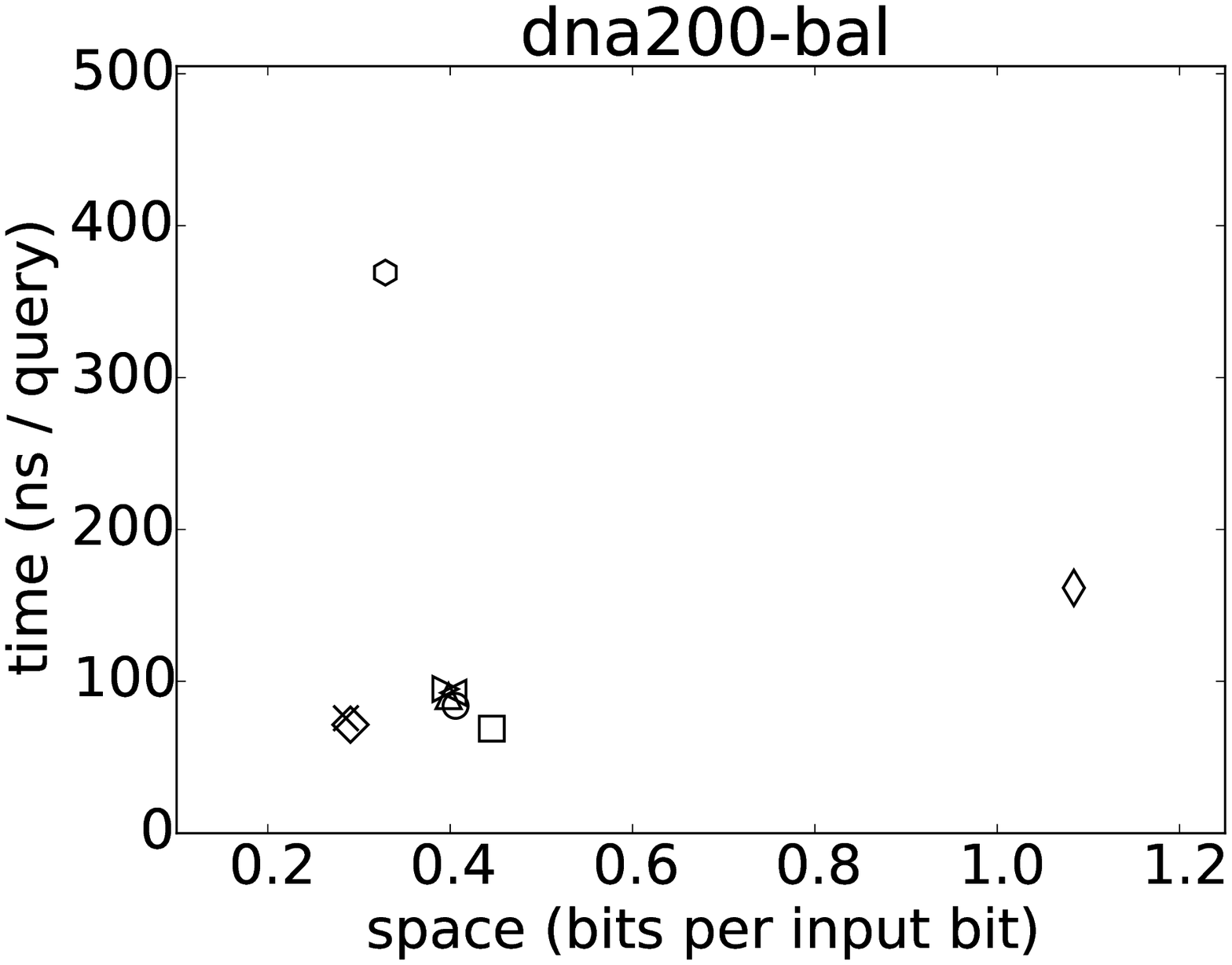}
\includegraphics[width=0.48\textwidth,scale=1.0]{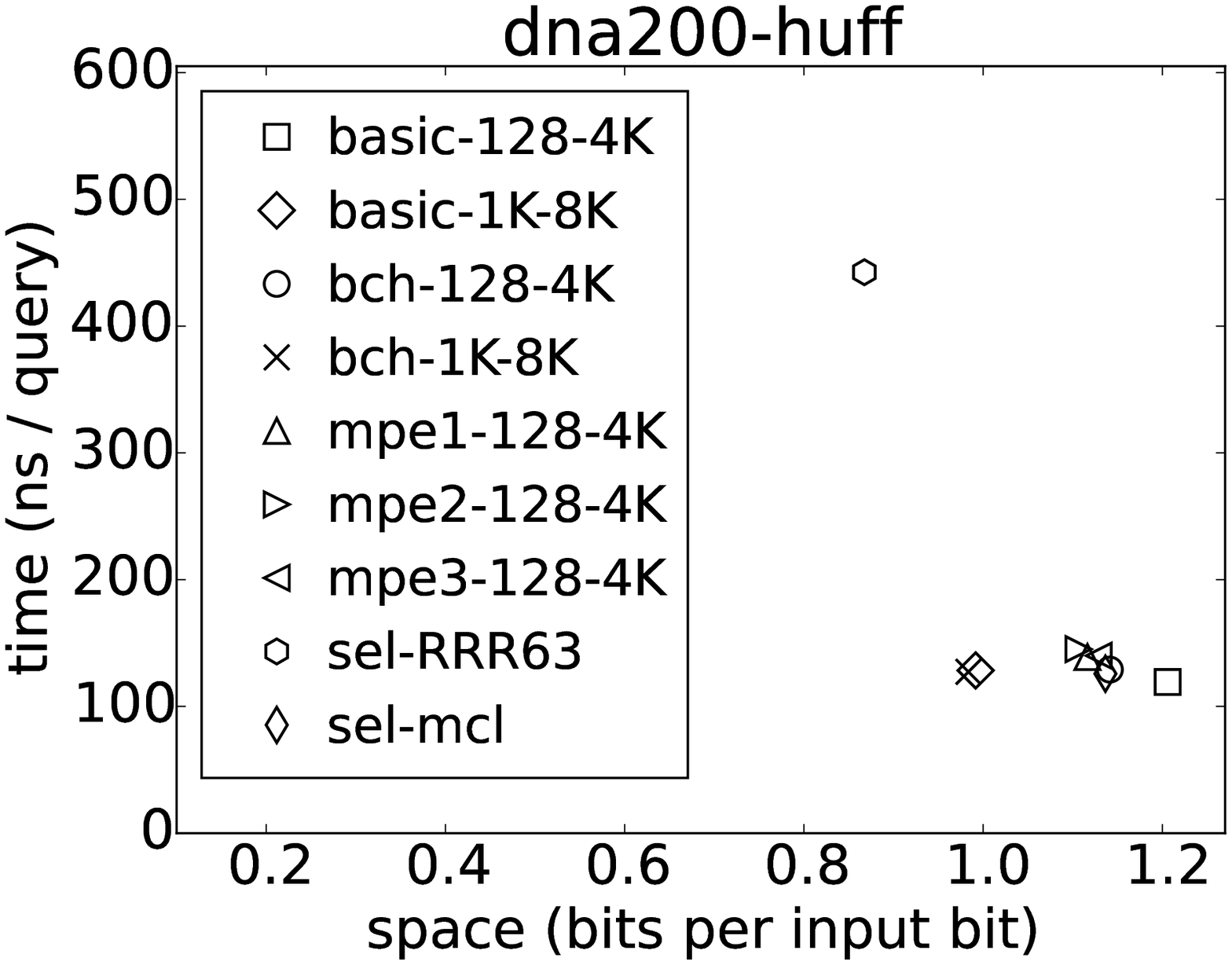}
}
\centerline{
\includegraphics[width=0.48\textwidth,scale=1.0]{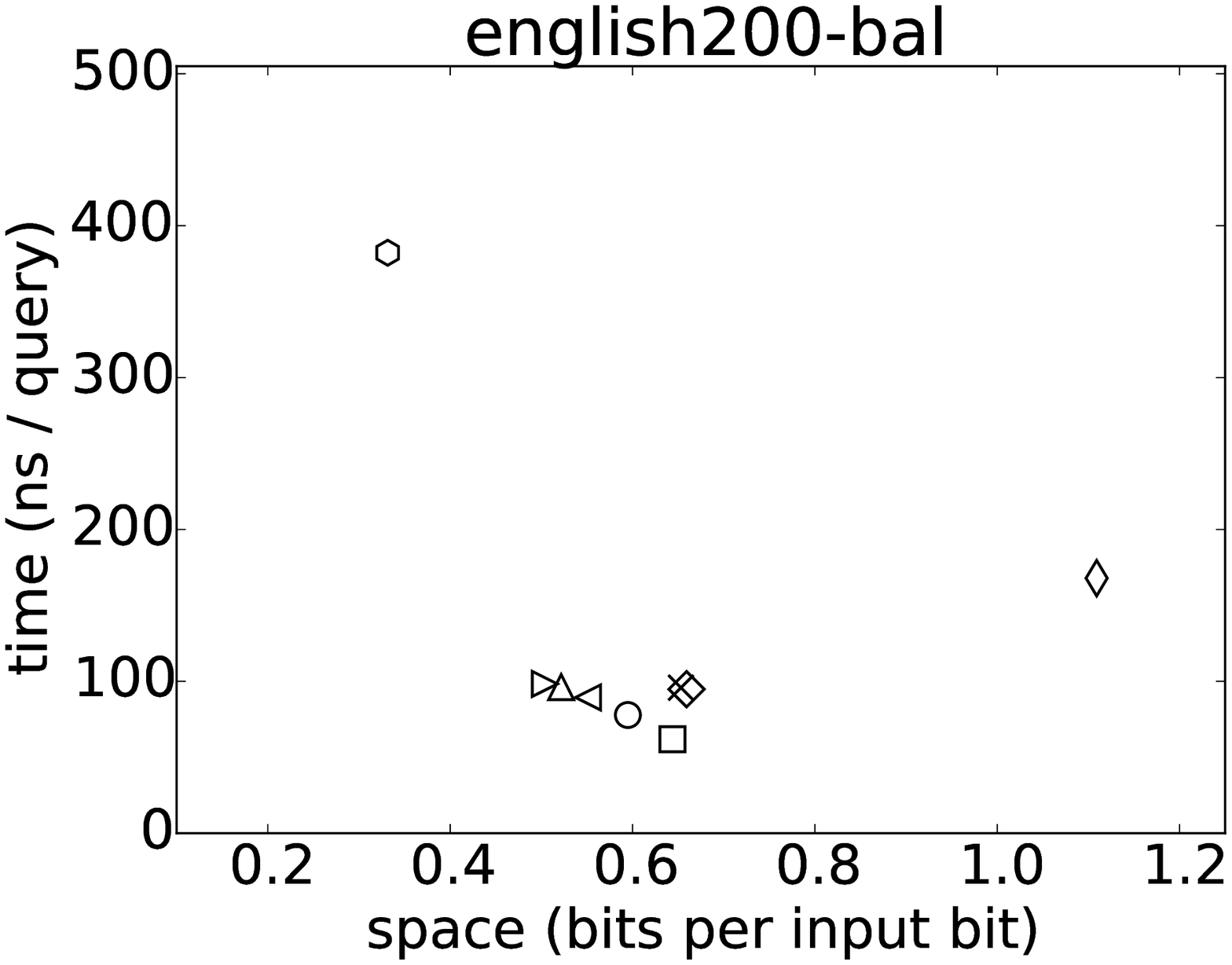}
\includegraphics[width=0.48\textwidth,scale=1.0]{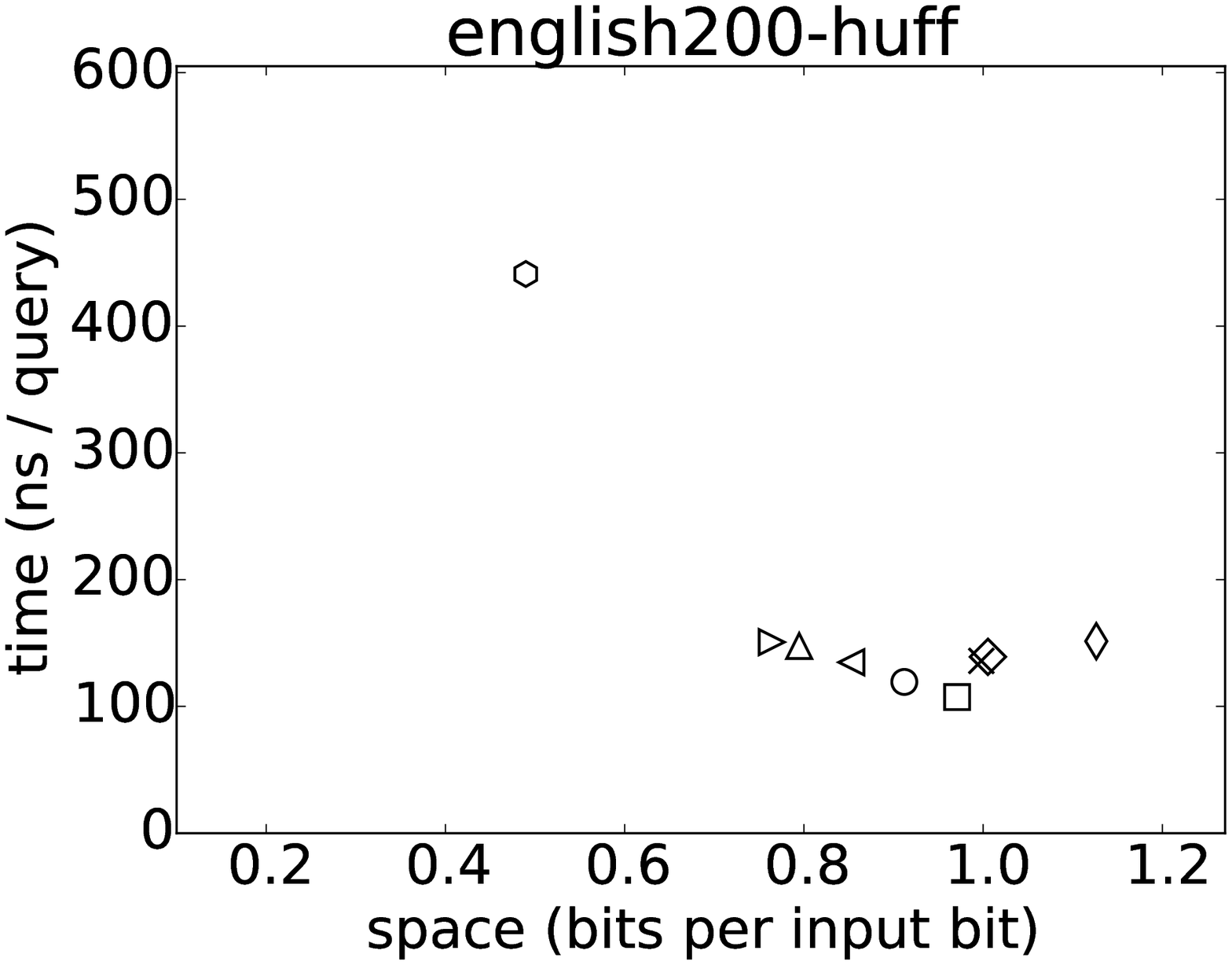}
}
\centerline{
\includegraphics[width=0.48\textwidth,scale=1.0]{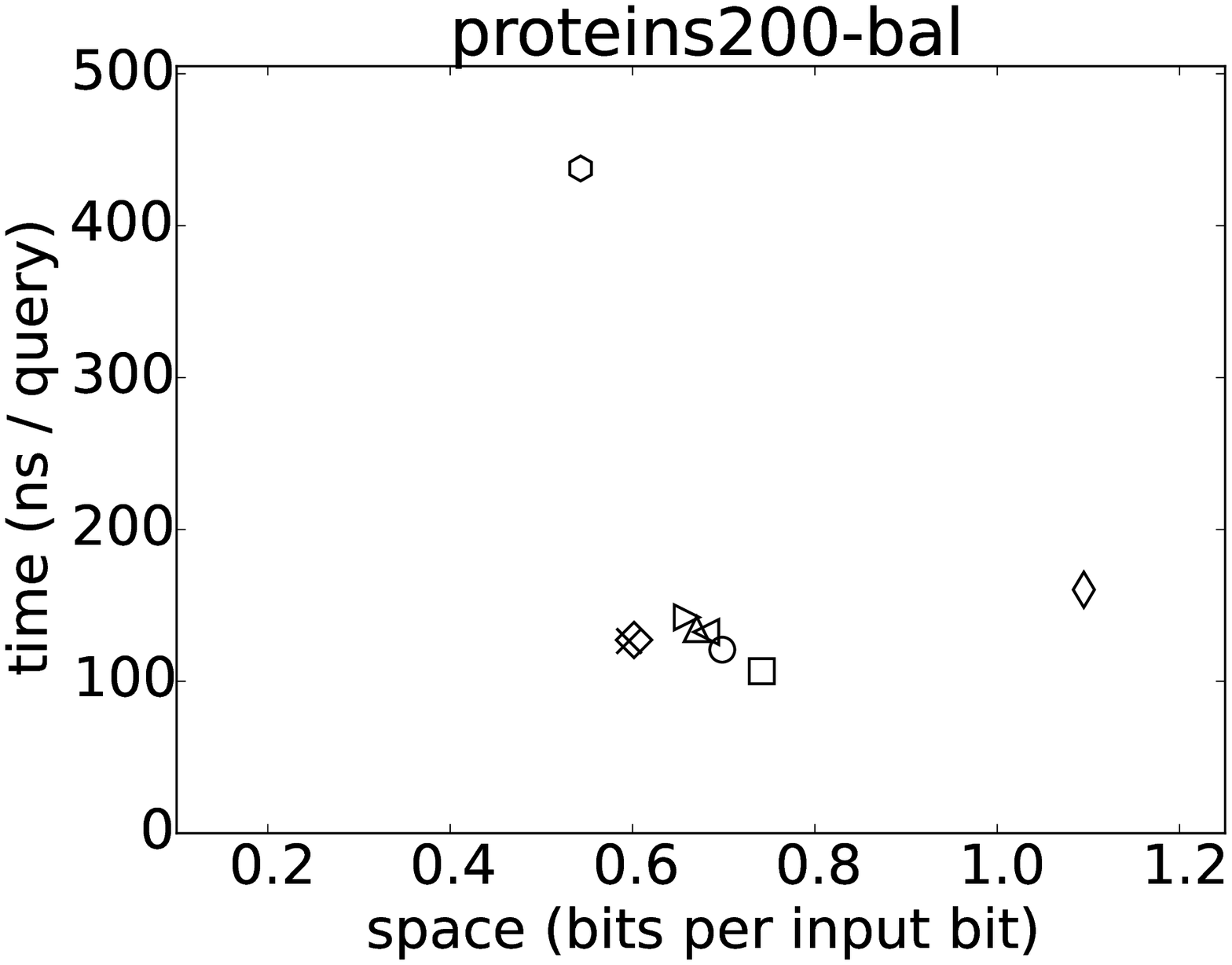}
\includegraphics[width=0.48\textwidth,scale=1.0]{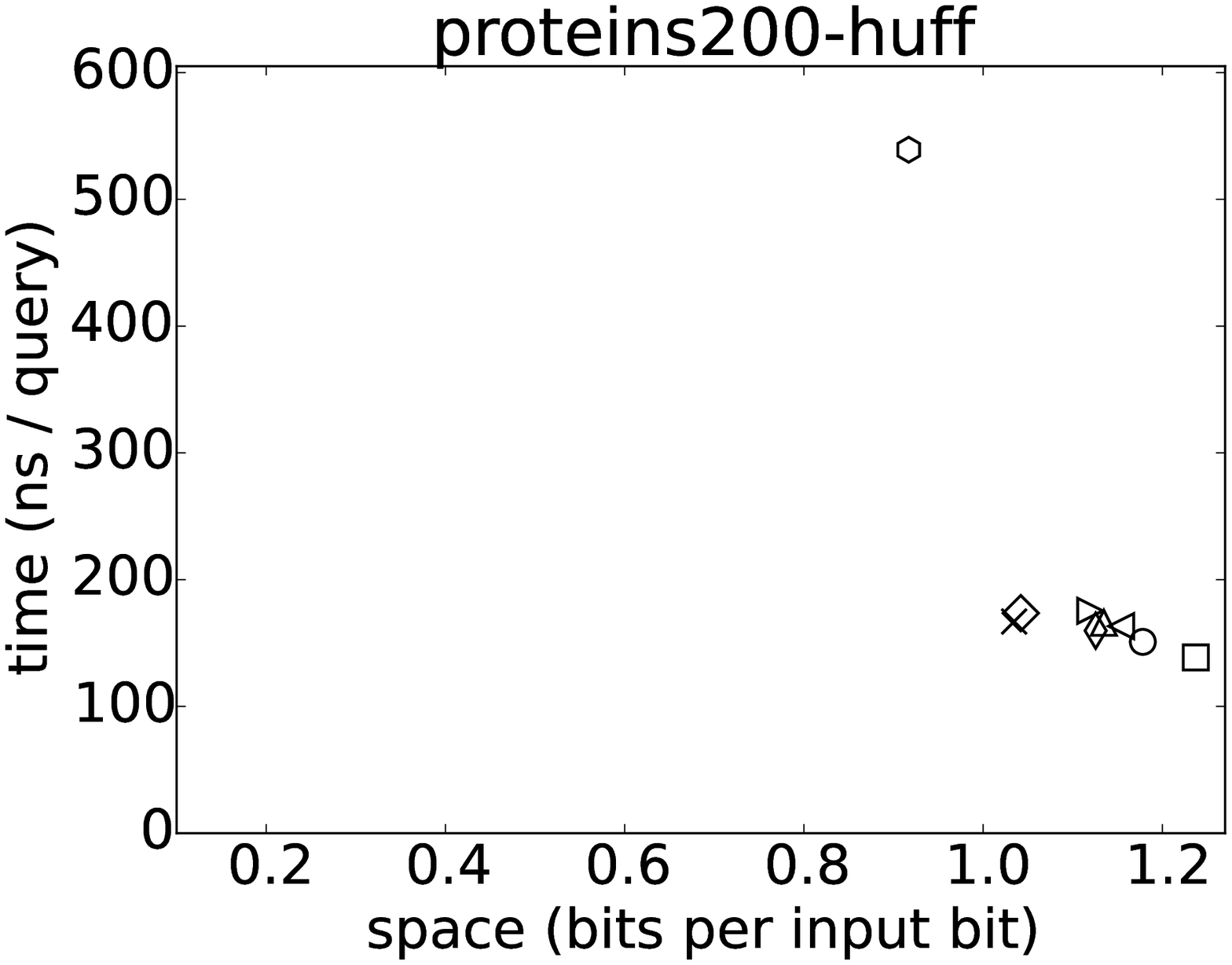}
}
\centerline{
\includegraphics[width=0.48\textwidth,scale=1.0]{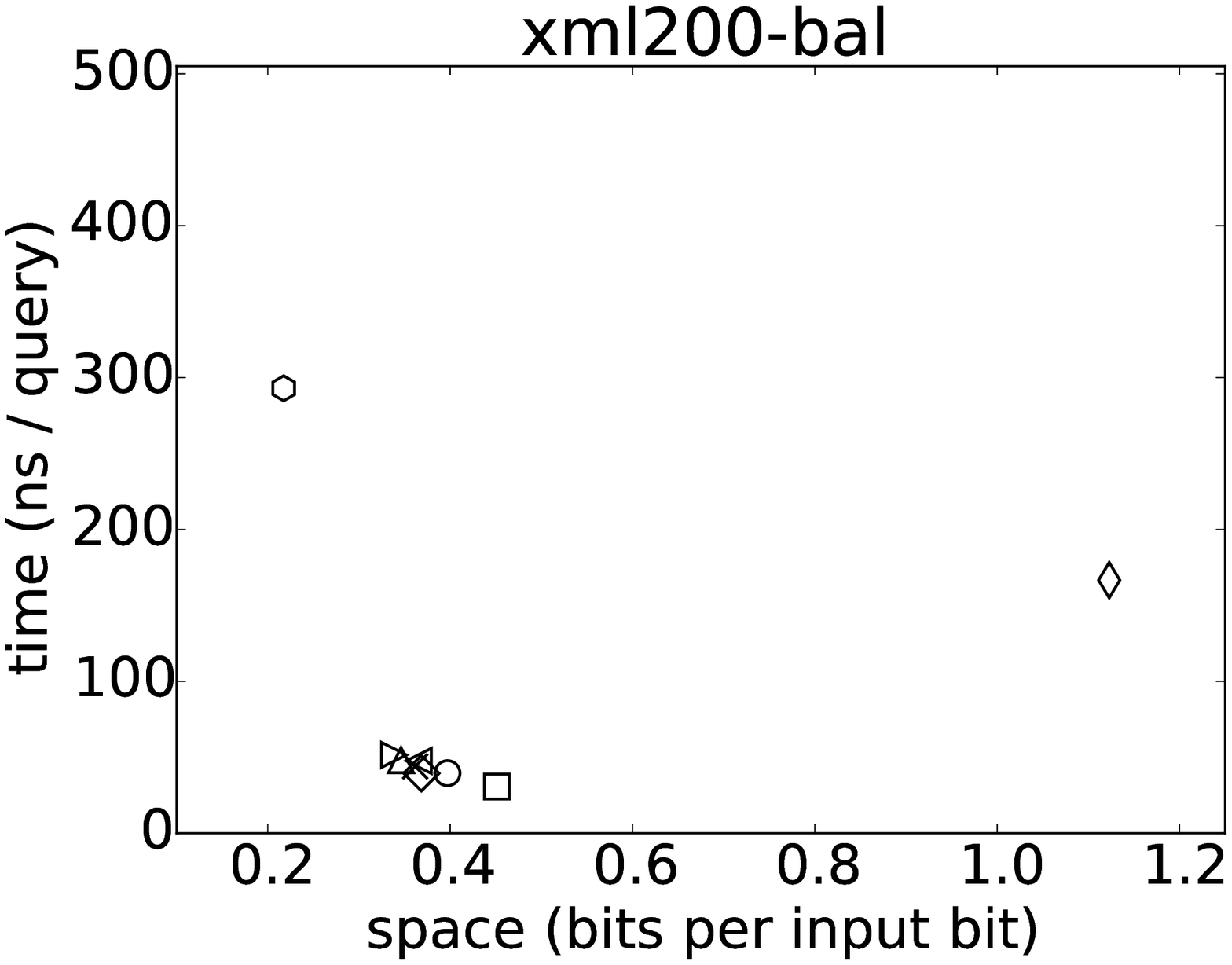}
\includegraphics[width=0.48\textwidth,scale=1.0]{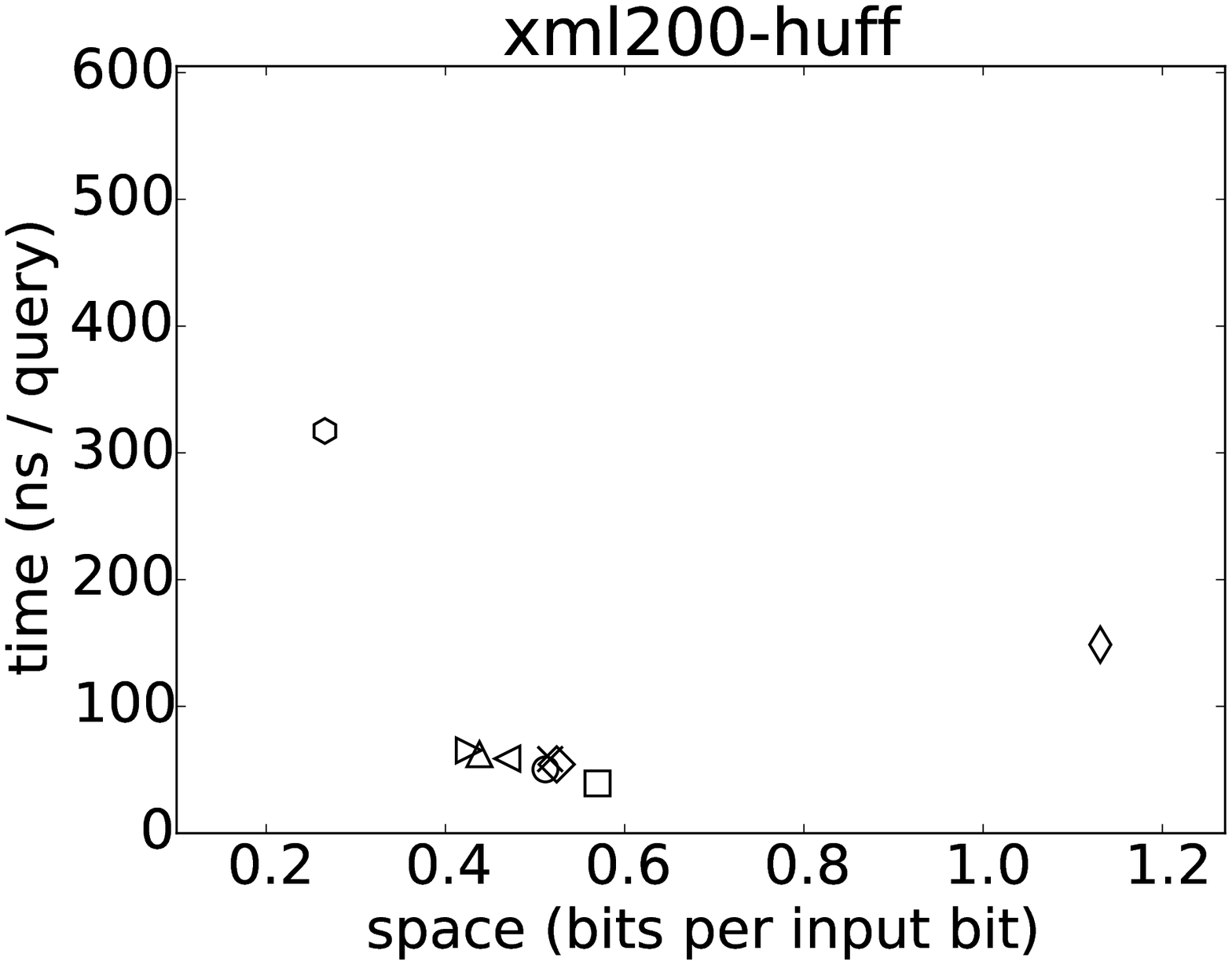}
}
\caption[Results]
{Select query times (in ns), averaged over 100M random queries, 
and used spaces for real data.
The datasets on the left: concatenated balanced WTs, 
on the right: concatenated Huffman-shaped WTs.}
\label{fig:cselects1}
\end{figure}

The proposed select variants (Fig.~\ref{fig:cselects1}) 
are about 3--4 times faster than {\em sel-RRR63} (and even 6--7 times faster 
on both XML datasets), yet offering worse compression. 
The numbers in the names of our variants are the used parameters, e.g.,
variant-128-4K denotes $\ell = 128$ and $thr = 4096$.
Compared to the non-compressed select variant, {\em sel-mcl}, 
whose space requirement is about $1.13n$ in all the cases, 
our solutions are both more compact and faster in six cases out of eight 
and obtain mixed results in the remaining two cases.

\begin{figure}
\centerline{
\includegraphics[width=0.48\textwidth,scale=1.0]{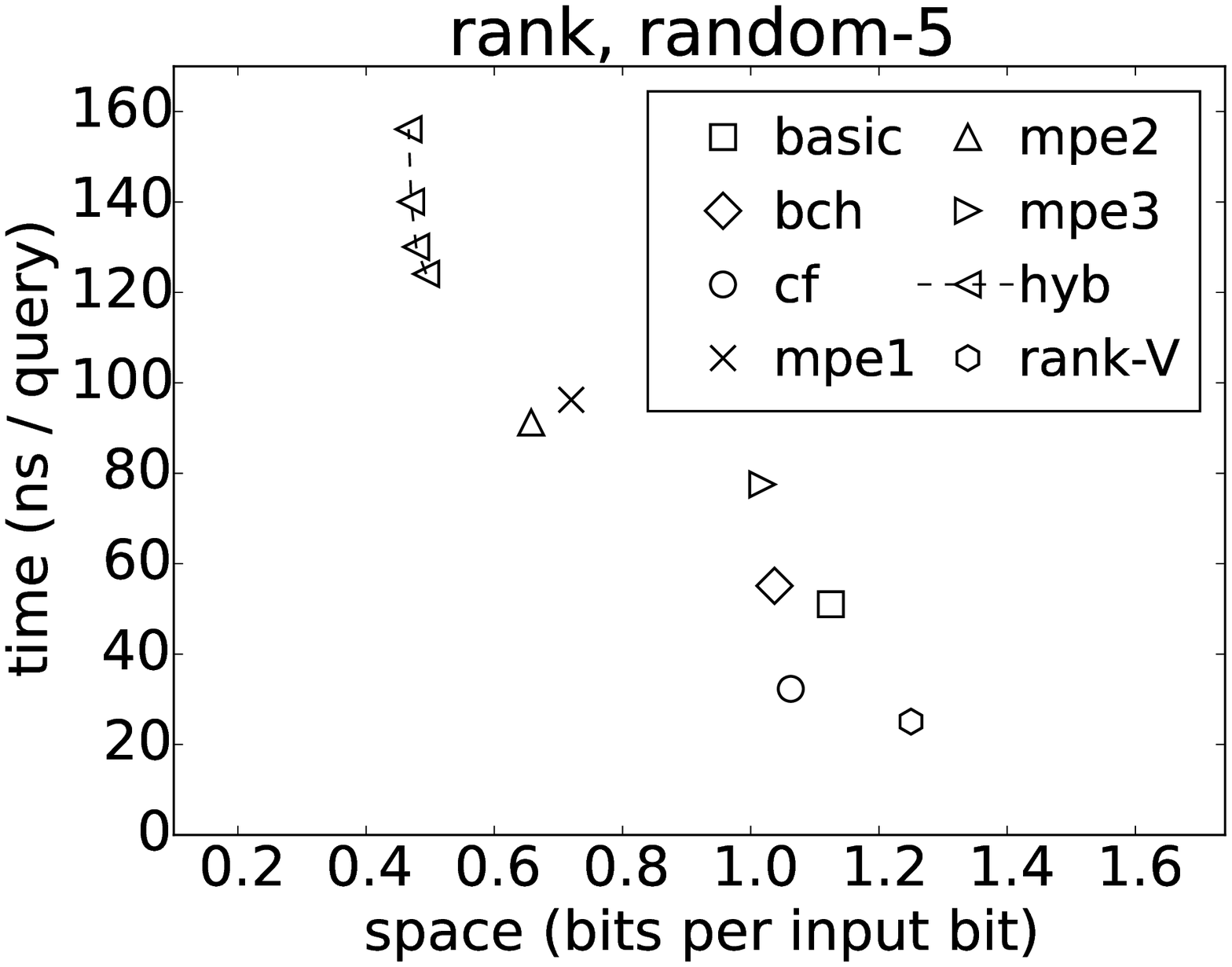}
\includegraphics[width=0.48\textwidth,scale=1.0]{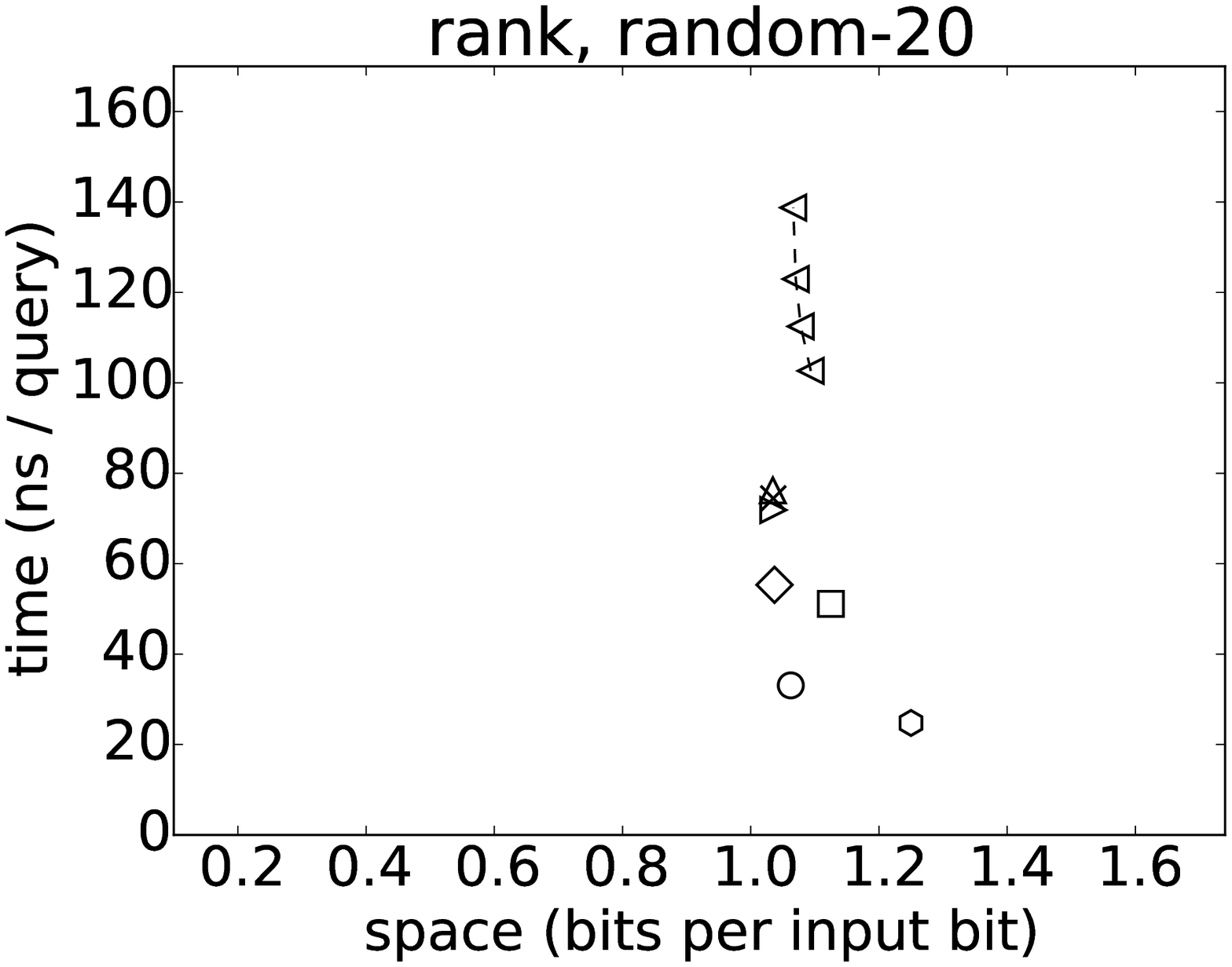}
}
\centerline{
\includegraphics[width=0.48\textwidth,scale=1.0]{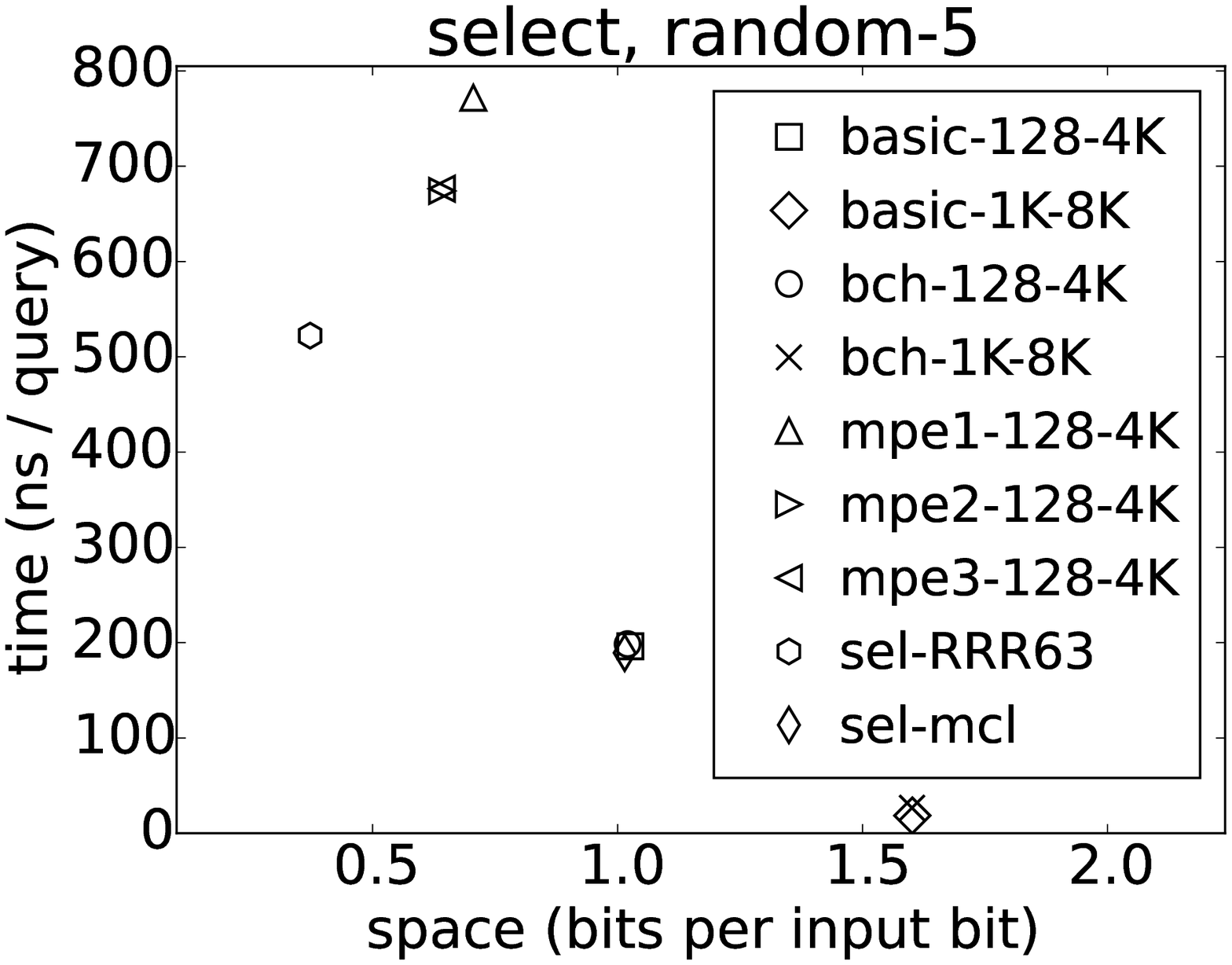}
\includegraphics[width=0.48\textwidth,scale=1.0]{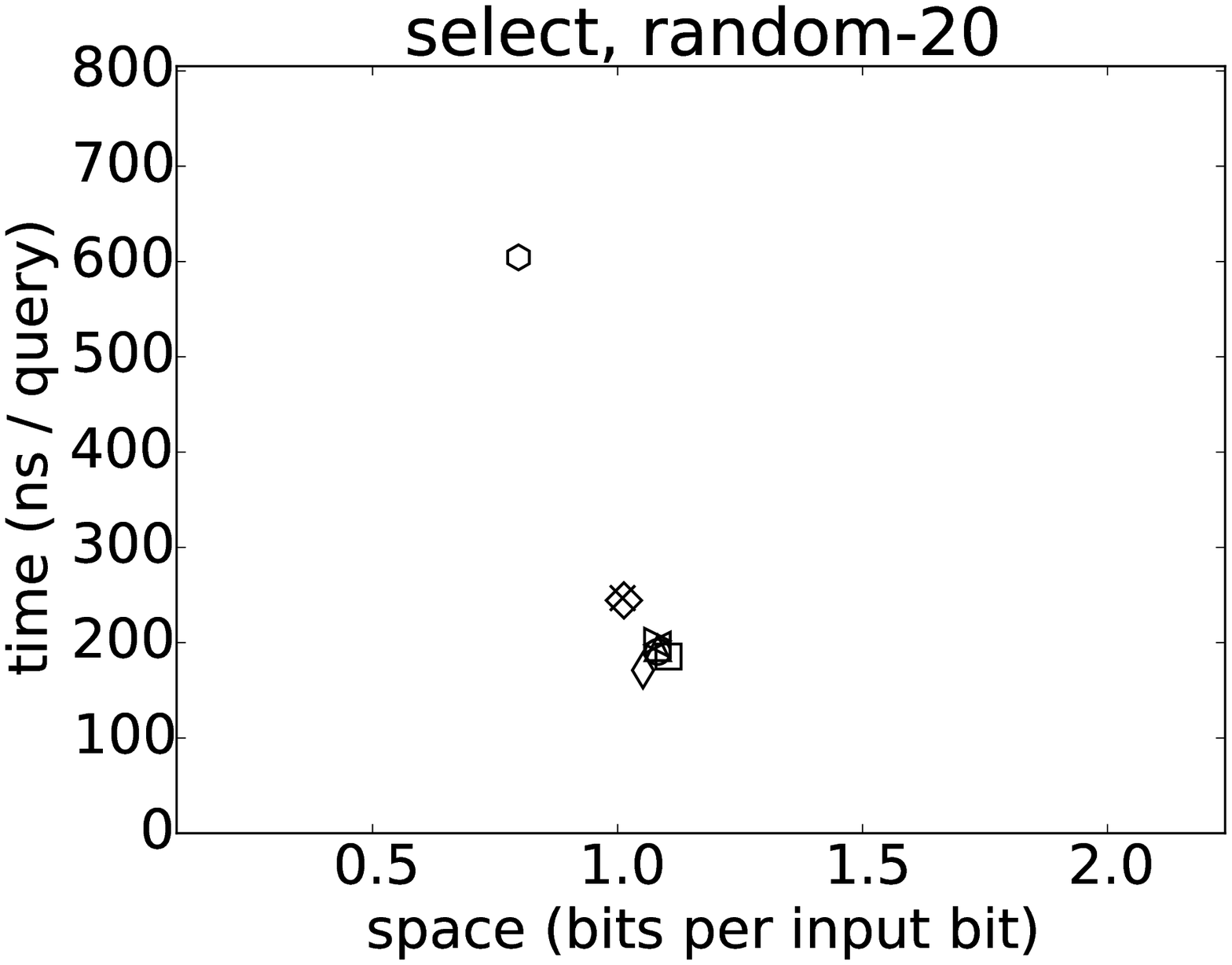}
}
\caption[Results]
{Rank (top) and select (bottom) query times (in ns), 
averaged over 100M random queries, and used spaces for random data 
with density of set bit 0.05 or 0.2.
The four points in a {\em hyb} series in the top figures 
correspond to parameters: 8, 16, 32 and 64.
}
\label{fig:cranks_cselects_random}
\end{figure}

Finally, we run the rank and select algorithms on 
uniformly random bit-vectors of size 200\,MB, 
with the density of bits 1 set to 5\% (random-5) or 20\% (random-20), 
as presented in Fig.~\ref{fig:cranks_cselects_random}.
Our rank results (the top row) in the 20\% case are quite competitive 
(especially the variant {\em cf}), but they are less 
attractive when the set bit density is lower.
The select query times (the bottom row)
vary a lot for the low density case, with major differences in space as well, 
but our solutions are Pareto-optimal only if 
practically all the blocks are sparse, which is obviously not interesting.
There is nothing to boast about the 20\% case either.

Overall, we can conclude that no single rank or select solution 
dominates over a large range of data characteristics.
For random data with low density, the rank {\em hyb}~\cite{KKP14} 
may be the winner. 
For bit-vectors with higher-order redundancies, 
the Beskers and Fischer~\cite{BF14} algorithms are the most competitive ones.
Finally, for real data from FM-indexes (concatenated binary wavelet trees) 
the solutions proposed in this paper tend to win, 
sometimes by a large margin.

\section{Conclusion}
\noindent
Rank and select are major components of many compressed data structures.
In spite of a great progress in their implementations in recent years, 
the ``ultimate'' variants have not yet been found, as this paper tries 
to demonstrate.
Not claiming big originality of our techniques, we have to point out 
that a number of the proposed variants represent new Pareto frontiers 
for query time and data structure size for real data.
One of our achievements is a (moderately) compressed select variant 
answering queries in about 100\,ns.
The key idea used in our algorithms is efficient handling of aligned runs 
of zeros and ones of specified size: either relatively large blocks 
(the technique used in all our variants) or 16-bit chunks 
(all the {\em mpe} variants).
On the other hand, reducing the number of memory accesses, 
a leitmotif in rank/select research for many years, 
still has some potential, as demonstrated by the {\em cf} variant.

Several aspects of our ideas require further research. 
The {\em cf} variant for the rank operation could be modified to handle 
compressed blocks.
The proposed select solutions work for only one binary digit (e.g., 1) 
and the current data structures would have to be doubled 
to handle both digits. 
Yet, it may be possible to modify them for data reuse, without a large 
drop in performance.
Finally, the rank variants have to be embedded in a full FM-index.
According to our preliminary experiments, an FM-index with Huffman-shaped
binary wavelet tree and the {\em rank-mpe2} variant performs the 
count query in time shorter by about 7--11\% than the FM-hybrid 
with the superblock size 8.
If the hybrid's superblock is increased to 64 (when the FM-index with {\em rank-mpe2} still wins in compression), the gap grows to 27--37\%.

\section*{Acknowledgement}
\noindent
The work was supported by the 
Polish National Science Centre under the project DEC-2013/09/B/ST6/03117 
(both authors).

\bibliographystyle{abbrv}
\bibliography{ranksel}

\end{document}